\documentclass[%
 reprint,
 amsmath,amssymb,
 aps,
prl
]{revtex4-2}

\usepackage{graphicx}
\usepackage{dcolumn}
\usepackage{bm}
\usepackage{dsfont} 

\newcommand{\llangle}{\langle\!\langle}
\newcommand{\rrangle}{\rangle\!\rangle}

\begin{document}

\preprint{APS/123-QED}

\title{Local connectivity balance shapes population dynamics in random recurrent networks}

\author{Shotaro Takasu}
    \email{stakasu@scripps.edu}
\author{Richard Gast}
\author{Ann Kennedy}
\affiliation{Department of Neuroscience, Scripps Research, San Diego, CA}

\date{\today}

\begin{abstract}
Disordered dynamical systems comprising many interacting units, from ecological communities to neural circuits, are ubiquitous, and understanding how connectivity shapes their collective behavior is a central theoretical challenge. One long-recognized feature of neural circuits is local connectivity balance, in which the excitatory and inhibitory weights converging onto each unit approximately cancel. Although local connectivity balance has been proposed to serve functions such as gating incoming signals, its effect on collective network dynamics remains unclear. Here we analytically study randomly connected recurrent networks with varying degrees of local connectivity balance. We show that this balance leaves the connectivity spectrum unchanged yet drastically reshapes the dynamics in a manner that depends critically on the single-unit nonlinearity. Local balance suppresses unbounded growth of the network state and stabilizes network dynamics when the activation function scales linearly or faster, whereas it drives the network into chaos when the activation function is sub-linear or saturating. Importantly, these effects vanish for odd activation functions, which are commonly assumed in previous work. We further find that, for saturating nonlinearities, the effective dimension of the dynamics varies nonmonotonically with the degree of balance. We show that all these phenomena arise from a unifying mechanism: the suppression of a self-generated feedback input by local connectivity balance. Our results identify local connectivity balance as a previously overlooked control parameter for collective dynamics in realistic disordered networks.
\end{abstract}

\maketitle

{\it Introduction ---}
Disordered dynamical systems of many interacting units appear across a remarkable range of fields, from neural circuits~\cite{amari1972,hopfield1982} and gene regulatory networks~\cite{kauffman1969} to ecological communities~\cite{volterra1928} and collective synchronization phenomena~\cite{kuramoto1975}. Despite their different microscopic interpretations, these systems share a common theoretical challenge: understanding how the statistical structure of disordered interactions shapes emergent collective dynamics. A simple and widely used framework for addressing this challenge is the random recurrent network~\cite{sompolinsky1988}, in which a large number of units interact through random connections. Although most commonly associated with neural circuits, variants of this framework have been used across all of these domains~\cite{sidhom2020,pham2024,hauck2026,pruser2024}.

The seminal work of Sompolinsky {\it et al.}~\cite{sompolinsky1988} showed that such a network undergoes a sharp transition from a quiescent fixed point to chaotic activity as the variance of the random couplings increases, and that this transition can be characterized exactly in the large-network limit using dynamical mean-field theory. This framework has since become a cornerstone for analyzing high-dimensional disordered dynamical systems. Building on this foundation, a large body of subsequent work has characterized how the network dynamics are reshaped by features such as the weight distribution~\cite{haruna2023, kusmierz2020, wardak2022, marti2018}, external input~\cite{molgedy1992, schuecker2018, toyoizumi2011, massar2013, rajan2010, engelken2022, takasu2024}, structured connectivity~\cite{mastrogiuseppe2018, kadmon2015, hayakawa2020, kusmierz2025, shao2025, clark2025, khajeh2022, landau2018}, heterogeneity across units~\cite{aljadeff2015, tomita2025}, the in- and out-degree distributions~\cite{metz2025}, and the single-unit dynamics~\cite{stern2014, keup2021, krishnamurthy2022}.

A structural property that has long been recognized is local connectivity balance, in which the inputs converging onto each unit sum to approximately zero. In neuroscience, this corresponds to {\it local synaptic balance}, the balancing of excitatory and inhibitory synaptic inputs at the single-neuron level. Experimental evidence for such balance spans multiple brain regions, including the visual~\cite{xue2014}, somatosensory~\cite{iascone2020}, and auditory cortices~\cite{wehr2003}, as well as hippocampal area CA3~\cite{atallah2009}. Computational studies have suggested several functional roles for local synaptic balance, including gating incoming signals~\cite{vogels2011} and enhancing robustness to noise~\cite{rubin2017}. 

In large disordered recurrent networks, local connectivity balance has been studied primarily through its effect on the eigenvalue spectrum of the connectivity matrix. In connectivity matrices that would otherwise exhibit outlier eigenvalues detached from the bulk, enforcing this balance removes the outliers~\cite{tao2013}, thereby suppressing the realization-to-realization fluctuations in the dynamics that such outliers induce~\cite{rajan2006, landau2018, hayakawa2020}. Beyond this spectral cleaning, however, whether and how local connectivity balance shapes the collective dynamics has remained largely unexplored.

In this letter, we report that the effects of local connectivity balance on collective dynamics emerge only when a common assumption in previous studies is relaxed: that the activation function is odd~\cite{sompolinsky1988,haruna2023, wardak2022,marti2018, molgedy1992, schuecker2018, toyoizumi2011,massar2013, takasu2024, mastrogiuseppe2018,hayakawa2020, kusmierz2025, shao2025, clark2025, aljadeff2015,schuessler2020, crisanti2018, clark2023, zhao2026}. Furthermore, we show that local connectivity balance has contrasting effects on dynamical stability: it suppresses unbounded growth of the network state (hereafter referred to as divergence) but induces chaos.


\begin{figure*}[t]
    \centering
    \includegraphics[width=\textwidth]{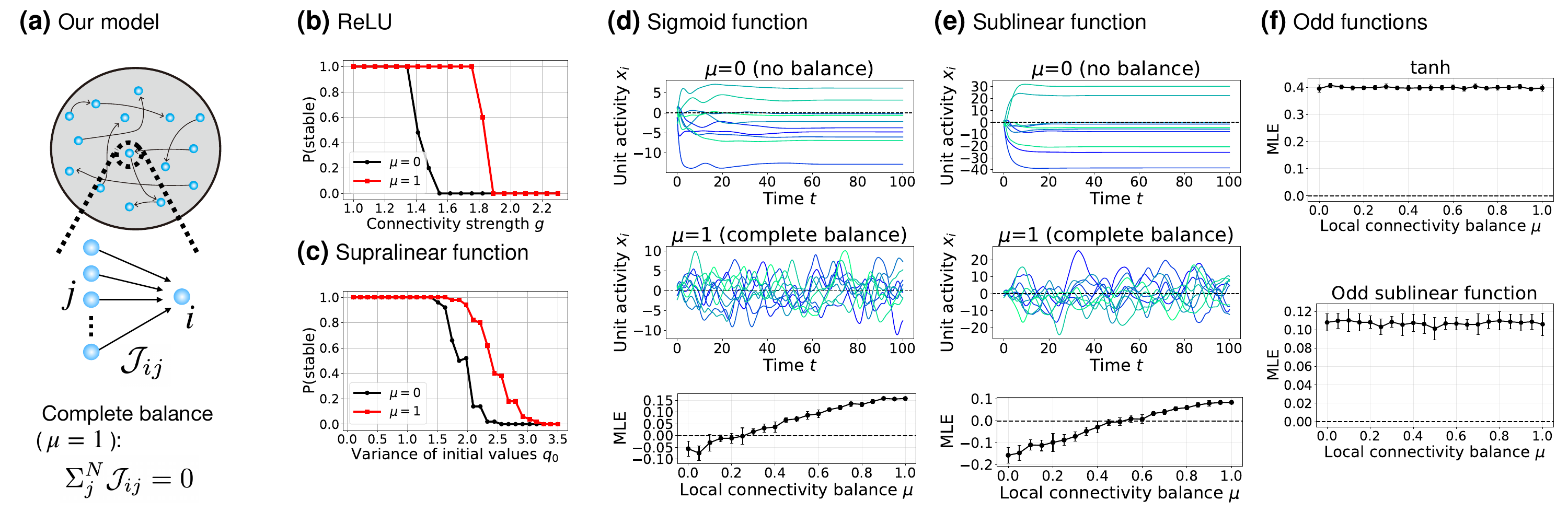}
    \caption{%
    Numerical simulations illustrating how local connectivity balance affects network dynamics. (a) Schematic of the random recurrent network model with local connectivity balance.
    (b,c) Fraction of stable (non-divergent) dynamics, $P(\mathrm{stable})$, for the ReLU function and the supra-linearly scaling function, respectively. For each parameter value, we simulate 50 networks with independent realizations of $\mathcal{J}$ and compute the fraction that converge to a fixed point. We use $g=1.0$ in (c), and a simulation length of $T=1000$.
    (d,e) Activities of 10 representative units for the sigmoid (d) and sub-linear (e) networks, at $\mu=0$ (top) and $\mu=1$ (middle). The bottom panels show the numerically computed maximum Lyapunov exponent (MLE) as a function of $\mu$; error bars indicate the mean $\pm$ std over 10 network realizations. We use $g=10$ for the sigmoid network and $g=5$ for the sub-linear network.
    (f) Numerically computed MLE for tanh (top) and the odd sub-linear function (bottom), both with $g=10$.
    In all simulations, the network size is $N=1000$.
    }
    \label{fig:examples}
\end{figure*}

{\it Model ---}
We study the population dynamics of a randomly connected network of $N \gg 1$ units~\cite{sompolinsky1988} described by
\begin{align} \label{eq:network dynamics}
    (1+\partial_t)x_i = \sum_{j=1}^N \mathcal{J}_{ij}\phi(x_j),
\end{align}
where $x_i$ and $\phi(x_i)$ are the preactivation and the activation of the microscopic units, respectively, and $\mathcal{J}$ denotes the recurrent connectivity. By varying the form of the single-unit dynamics (left-hand side) and unit interactions (right-hand side), this network equation can describe a broad range of systems, such as the generalized Lotka-Volterra model of ecosystems~\cite{sidhom2020}, gene regulatory networks~\cite{pham2024, hauck2026}, the Kuramoto model of synchronization phenomena~\cite{pruser2024}, opinion dynamics~\cite{baumann2020}, and game theory~\cite{galla2013}. A discrete-time version of our model with connectivity that is drawn independently at each time step is equivalent to a deep feedforward neural network~\cite{poole2016, schoenholz2017}. Thus the theory we develop here applies to diverse domains.

We define the recurrent connectivity as
\begin{align} \label{eq:connectivity}
    \mathcal{J}_{ij} := J_{ij} - \frac{\mu}{N}\sum_{k=1}^N J_{ik}\hspace{10pt} \left( 0 \leq \mu \leq 1\right),
\end{align}
where $J_{ij}$ is drawn i.i.d.\ from a Gaussian distribution with mean $0$ and variance $g^2/N$. The parameter $\mu$ controls the degree of local connectivity balance: noting that $\sum_{j=1}^N \mathcal{J}_{ij} = (1-\mu)\sum_{j=1}^N J_{ij}   $, we see that the sum of incoming weights onto a unit approaches zero as $\mu$ approaches 1 (Fig.~\ref{fig:examples}(a)). We thus refer to $\mu=1$ as {\it complete balance}~\cite{rajan2006}, and $\mu=0$  as {\it no balance}. The latter, combined with an odd sigmoid activation function, reduces to the conventional random network model of~\cite{sompolinsky1988, molgedy1992,toyoizumi2011, schuecker2018} and others.

Local connectivity balance has a vanishing impact on the weight distribution and eigenvalue spectrum of $\mathcal{J}$. In the limit $N\to \infty$, we see that $\mathcal{J}_{ij}\to J_{ij}$ because $J_{ij}$ scales as $\mathcal{O}(1/\sqrt{N})$ while the balance term is of order $\mathcal{O}(1/N)$. It is conceivable that the balance term might still produce eigenvalues of $\mathcal{J}$ that fall outside the bulk spectrum of $J$, because it acts as a rank-1 perturbation to $J$~\cite{mastrogiuseppe2018, schuessler2020}. However, we prove that the magnitude of the balance term is small enough that no outlier eigenvalues are created (see Supplemental Material~\cite{SM} for a proof). Therefore, $\mathcal{J}$ cannot be distinguished from a Gaussian random matrix solely based on its weight distribution or eigenvalue spectrum. Nevertheless, we show in numerical simulations below that local connectivity balance has a strong, systematic effect on network dynamics.

{\it Numerical simulations ---}
The choice of activation function strictly determines the possible dynamical states of network models. For linearly and supra-linearly scaling functions ($\phi(x) \sim x^{\alpha}$ with $\alpha\geq 1$ for $x\gg 1$), network dynamics either converge to a fixed point or diverge. Conversely, for sub-linearly scaling functions ($\phi(x) \sim x^{\alpha}$ with $0< \alpha < 1$ for $x\gg 1$) and bounded functions such as the sigmoid function, dynamics are either fixed-point or chaotic. For each activation function family, we examine the effect of local connectivity balance on the transition between dynamical regimes.

First, we consider networks with linear and supra-linear activation functions, beginning with the rectified linear (ReLU) function $\phi(x)=\max(0,x)$. In  numerical simulations of network dynamics, we find that the network activity of ReLU networks with no balance diverges once $g$ exceeds about $1.4$ (Fig.~\ref{fig:examples}(b), black line). In contrast, ReLU networks with complete balance remain stable up to $g\approx 1.8$, indicating that local connectivity balance suppresses explosive network dynamics (Fig.~\ref{fig:examples}(b), red line).

We next consider the supra-linearly scaling function $\phi(x) = \left( \max(0,x) \right)^2$, and again find local connectivity balance to have a stabilizing effect. Here the zero fixed point is linearly stable regardless of $g$, and whether the network state converges to zero depends on its initial state. In Fig.~\ref{fig:examples}(c), we again run numerical simulations of network dynamics, with the initial state assigned randomly according to $x_i(0) \overset{\rm i.i.d.}{\sim} \mathcal{N}(0, q_0)$. As in the ReLU case, local connectivity balance markedly increases the fraction of network states that converge to zero as a function of $q_0$. Thus, for linearly and supra-linearly scaling activation functions, we find that local connectivity balance suppresses divergence and thus stabilizes the network dynamics.

In contrast, we find that local connectivity balance has a destabilizing effect on networks with bounded or sub-linearly scaling activation functions. Figs.~\ref{fig:examples}(d,e) show simulations of networks with $\phi(x)=1/(1+e^{-x})$ (sigmoid) and $\phi(x)=\sqrt{(x+1)+\sqrt{(x+1)^2+3}}$ (a sub-linearly scaling function used in~\cite{schmidt2018}), showing that local connectivity balance destabilizes the fixed-point dynamics and drives the network into chaos. We numerically compute the maximum Lyapunov exponents (MLE)~\cite{Pikovsky2016} of both systems, and find that they increase monotonically with $\mu$ (bottom figures in Fig.~\ref{fig:examples}(d,e)).

Interestingly, we find that local connectivity balance has no impact on the stability of the network dynamics for odd activation functions. The MLE of these networks is constant with respect to $\mu$, regardless of whether the activation function is saturating ($\phi(x)=\tanh(x)$) or non-saturating but sub-linearly scaling (e.g. $\phi(x)=x/(1+x^2)^{1/4}$) (Fig.~\ref{fig:examples}(f)). In the text below, we analytically show that these networks' dynamics are unaffected by local connectivity balance in the limit $N\to \infty$.

Taken together, we describe a novel relationship between the effect of local connectivity balance on network dynamics and the unit activation function. While connectivity balance has no impact on network dynamics for odd activation functions, it otherwise significantly influences the dynamical stability, suppressing  divergence for linearly and supra-linearly scaling functions and inducing chaos for sigmoid and sub-linearly scaling functions.

{\it Unifying mechanism ---}
These seemingly contrasting effects of local connectivity balance can be understood intuitively by decomposing the network dynamics as
\begin{align} \label{eq:decomposition}
    (1+\partial_t)x_i = \sum_{j=1}^N J_{ij}\left( \phi(x_j)-\langle \phi \rangle\right) + (1-\mu)\langle \phi \rangle v_i,
\end{align}
where $\langle \phi \rangle:= (1/N)\sum_{k=1}^N \phi(x_k)$ is the population average of the unit activations and $v_i := \sum_{k=1}^N J_{ik}$ is the sum of incoming weights onto the $i$-th unit. $\langle \phi \rangle$ is typically time-independent unless the network state diverges, with the exception of coherent chaos, where it exhibits chaotic fluctuations even in the large-$N$ limit~\cite{landau2018, hayakawa2020}.

When $\phi$ is an odd function, $\langle \phi \rangle$ vanishes in the large-$N$ limit. Indeed, the central limit theorem, together with $\langle J_{ij}\rangle=0$, implies that the $x_i$ are distributed as a zero-mean Gaussian, and thus, $\langle\phi\rangle=0$ in this limit. This eliminates the $\mu$-dependence of Eq.~(\ref{eq:decomposition}), explaining why odd functions are insensitive to local connectivity balance.

Conversely, $\langle \phi \rangle$ is typically nonzero for non-odd activation functions like those in Figs.~\ref{fig:examples}(b-e). In this case, the second term in Eq.~(\ref{eq:decomposition}) can be thought of as a self-generated input scaled by a nonnegative factor $1-\mu$, which is fed back into the network through the weights $v_i$. In networks with dynamics that can potentially diverge, the self-generated input term contributes to divergence through positive feedback, and introducing local connectivity balance (decreasing $1-\mu$) weakens this feedback to prevent divergence. In contrast, it is known that in networks whose dynamics can exhibit chaos, external input suppresses the transition to chaos~\cite{molgedy1992,schuecker2018, toyoizumi2011,massar2013, rajan2010, engelken2022, takasu2024}. Therefore in these systems, self-generated input normally suppresses chaos, and introducing local connectivity balance removes this input and releases the suppressed chaos.

{\it Theoretical analysis ---}
The network dynamics of our model can be analyzed theoretically using dynamical mean-field theory~\cite{sompolinsky1988, dahmen2020}. In the limit $N\to \infty$, all units in the network become mutually independent, evolving according to
\begin{align} \label{eq:effective dynamics}
    (1+\partial_t) x = \eta,
\end{align}
where $\eta$ is a Gaussian process with mean $0$ and covariance function $C(t,s)$. The covariance function is determined self-consistently by
\begin{align} \label{eq:covariance function}
    C(t,s) = g^2 \left\{ \llangle \phi(t),\phi(s) \rrangle_\eta +(1-\mu)^2 \langle \phi(t) \rangle_\eta \langle \phi(s) \rangle_\eta \right\},
\end{align}
where $\langle \cdot \rangle_\eta$ and $\llangle \cdot \rrangle_\eta$ denote the first and second cumulants of $\phi$ with respect to the Gaussian process $\eta$, and we use the shorthand $\phi(t):= \phi(x(t))$. These mean-field equations can be rigorously derived via a path-integral formalism~\cite{crisanti2018, dahmen2020} (see Supplemental Material~\cite{SM} for a detailed derivation).

\begin{figure}[t]
    \centering
    \includegraphics[width=0.5\textwidth]{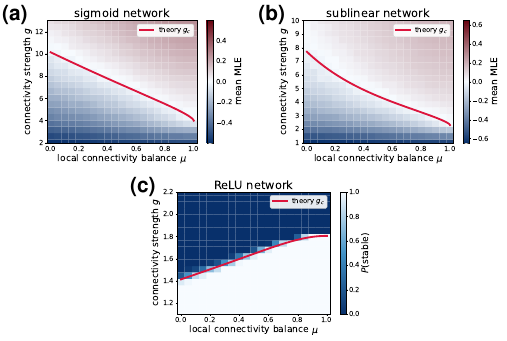}
    \caption{%
    Phase diagrams with analytically derived transition boundaries. (a,b) Phase diagrams for the sigmoid and sub-linear networks. The color scale represents the maximum Lyapunov exponent (MLE), averaged over 10 network realizations, and the red curves indicate the analytically derived critical connectivity strength $g_c$. (c) Phase diagram for the ReLU network. The color scale represents the fraction of stable realizations among 50 network realizations, and the red curve indicates $g_c$. The network size is $N=3000$ in all panels.
    }
    \label{fig:analytical solutions}
\end{figure}

First, we analyze the stability of the fixed point in networks with bounded or sub-linear activation functions, where chaos can emerge. Here we make three additional assumptions about the activation function: (1) it is increasing, (2) $\lim_{x\to \pm \infty}\phi'(x)=0$, and (3) $\phi'(x)$ is unimodal with its peak at $x=0$. The second assumption means that $\phi(x)$ scales sub-linearly or it is saturating, which prevents divergence and instead yields chaotic dynamics. These three assumptions are satisfied by a broad range of activation functions, including those used in Figs.~\ref{fig:examples}(d-f).

Assuming that the network state is at a fixed point, the preactivations $\{ x_i^{0} \}_{i=1}^N$ follow a Gaussian distribution with mean $0$ and variance $q$, which is determined self-consistently by $q=f(q)$, where
\begin{align} \label{eq:self-consistent equation for fixed point}
    f(q) :=g^2 \left\{  \llangle \phi,\phi \rrangle_{x\sim\mathcal{N}(0,q)} +(1-\mu)^2 \langle \phi \rangle_{x\sim\mathcal{N}(0,q)}^2\right\}.
\end{align}
This self-consistent equation follows by setting $s=t$ in Eq.(\ref{eq:covariance function}) and noting that, at a fixed point, Eq.(\ref{eq:effective dynamics}) gives $x_i^0=\eta_i^0$, so that $C(t,t)=q$. 

Fixed points become destabilized when the spectral radius $\rho$ of the Jacobian matrix exceeds one; for our system, $\rho=g\sqrt{\langle \phi'(x^0)^2 \rangle_{x^0 \sim \mathcal{N}(0,q)}}$. Since the eigenvalues of our Jacobian matrix are distributed over a disk-shaped bulk~\cite{ahmadian2015}, once $\rho$ exceeds one there are $\mathcal{O}(N)$ unstable directions, rendering the network dynamics chaotic for large $N$. The critical point for the transition  to chaos is therefore given by $g_c = 1/\sqrt{\langle \phi'(x^0)^2 \rangle_{x^0 \sim \mathcal{N}(0,q)}}$. Under our three assumptions on $\phi$, the solution of $q=f(q)$ is a decreasing function of $\mu$, and $\langle \phi'(x^0)^2 \rangle_{x^0 \sim \mathcal{N}(0,q)}$ increases as $q$ decreases. As a result, the spectral radius itself increases with $\mu$. We therefore see that local connectivity balance destabilizes the network dynamics by increasing the spectral radius (Fig.~\ref{fig:analytical solutions}(a,b)). See Supplemental Material~\cite{SM} for details of the theoretical analysis.

We now turn to the stability of the zero fixed point in ReLU networks. Unlike the sigmoid and supra-linear activation functions, the ReLU function is non-differentiable at zero. Consequently, the Jacobian of the network dynamics is undefined at $\bm{x}=\bm{0}$, precluding the Jacobian-based linear stability analysis. Instead, we determine the critical connectivity strength, $g_c$, from the dynamical mean-field equations (Eqs.(\ref{eq:effective dynamics}) and (\ref{eq:covariance function})). Under the assumption of stationarity, these equations yield
\begin{align} \label{eq: stationary dmft}
    (1-\partial_\tau^2)\Delta(\tau) = C(\tau),
\end{align}
where $\Delta(\tau):= \langle x(t)x(t+\tau) \rangle_\eta$ and $C(\tau):= C(t,t+\tau)$. At the critical point $g_c$, the network activity neither decays to zero nor diverges, justifying the following boundary conditions:
\begin{align} \label{eq: boundary condition}
    \Delta(0)=1,\ \Delta'(0)=0,\ \lim_{\tau\to\infty}\Delta'(\tau)=0.
\end{align}
Here, the first condition selects a nontrivial solution, whose overall amplitude is arbitrary owing to the scale invariance of the ReLU function, $\phi(ax)=a\phi(x)$ for $a\geq0$. Solving the dynamical mean-field equations subject to these boundary conditions yields
\begin{align}
    g_c = \left[ \frac{2}{1-\Delta(\infty)^2} \int_{\Delta(\infty)}^{1} h(\Delta, \mu) d\Delta \right]^{-\frac{1}{2}}, 
\end{align}
where 
\begin{align}
    h(\Delta, \mu) := \frac{1}{2\pi}\left[ \sqrt{1-\Delta^2} + \Delta(\pi-\arccos \Delta) -\mu(2-\mu) \right].
\end{align}
The value of $\Delta(\infty)$ is determined implicitly by
\begin{align}
    (1-\Delta(\infty)^2)h(\Delta(\infty), \mu) = 2\Delta(\infty)  \int_{\Delta(\infty)}^{1} h(\Delta, \mu) d\Delta.
\end{align}
As shown in Fig.~\ref{fig:analytical solutions}(c), the resulting analytical prediction for $g_c$ closely matches the transition boundary observed in numerical simulations and shows that $g_c$ increases monotonically with $\mu$. See Supplemental Material~\cite{SM} for a detailed derivation.

\begin{figure}[t]
    \centering
    \includegraphics[width=0.48\textwidth]{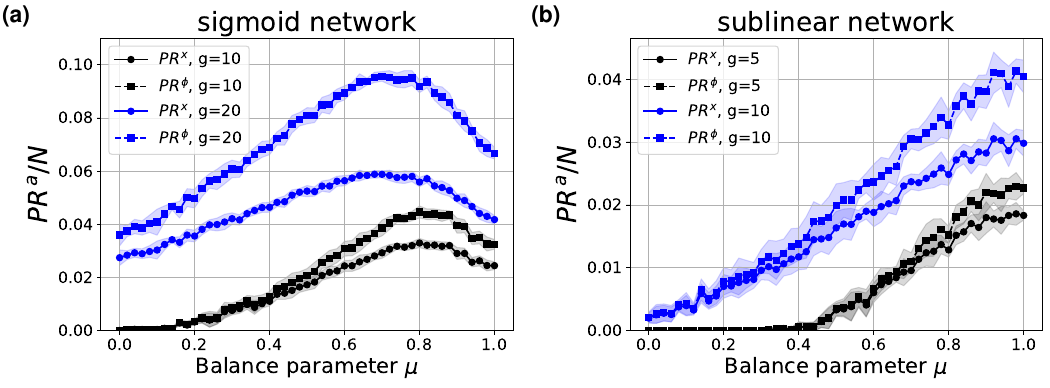}
    \caption{%
        Participation ratio $PR^a$ ($a\in\{x,\phi\}$, for the preactivation $x$ and the activation $\phi$) for (a) sigmoid and (b) sub-linear networks, shown for two values of $g$ each. Lines and shading indicate the mean and std over 10 network realizations. Network size is $N=3000$ and simulation time is $T=10^5$.
    }
    \label{fig:effective dimension}
\end{figure}

{\it Effective dimension ---}
Given that local connectivity balance monotonically increases the strength of the chaotic dynamics (as quantified by the MLE, Fig.~\ref{fig:examples}(d,e)), it is natural to expect that the effective dimension of the network population dynamics also increases monotonically with $\mu$. For the sigmoid network, however, this expectation fails: the participation ratio (PR) reaches its peak at an intermediate value of $\mu$ (Fig.~\ref{fig:effective dimension}(a)). Here, the PR, our measure of effective dimension, is defined as $PR^a= \frac{\left(\sum_{i=1}^N \lambda_i^a \right)^2}{\sum_{i=1}^N (\lambda_i^a)^2}$~\cite{rajan2010b}, where $\lambda_i^a$ is the $i$-th eigenvalue of the covariance matrix of the variables $a_i\in \{ x_i, \phi_i\}$, i.e., $\Sigma_{ij}^a = \llangle a_i(t), a_j(t)\rrangle_t$~\cite{clark2023}. This nonmonotonic behavior of the PR is robust over a wide range of $g$. In contrast, the sub-linear network shows a monotonic increase in the PR (Fig.~\ref{fig:effective dimension}(b)). Therefore, whether the PR increases as the local connectivity balance deviates from complete balance depends on the shape of the activation function.

The nonmonotonic behavior of the PR can be explained as follows. When $\mu=1$, each unit fluctuates around $x_i=0$ over time, because the time-averaged activity, $\langle x_i \rangle_t$, is exactly zero as long as the activation function satisfies the assumptions stated above (see Supplemental Material~\cite{SM} for a proof). Given that $\frac{1}{N} PR^a \ll 1$ at $\mu=1$ for both activation functions, the neural activity resides in an elongated ellipsoid in the $N$-dimensional state space. As noted above, when $\mu \neq 1$ the network receives a self-generated input, $(1-\mu)\langle \phi \rangle$, through the weights $v_i$, where $v_i = \sum_{k=1}^N J_{ik} \sim \mathcal{N}(0,g^2)$ for large $N$. As shown in a recent preprint~\cite{zhao2026}, a relatively small external input shifts the center of activity, $\langle \bm{x} \rangle_t $, along the high-variance directions of the spontaneous dynamics, analogous to the fluctuation-dissipation relation~\cite{kubo1957}. Therefore, if $\mu$ is slightly below $1$, the center of activity deviates significantly from zero in the units that contribute most to the high-variance directions. Since $\phi'(x)$ decreases as $x$ deviates from zero, the fluctuations along the high-variance directions decrease, which makes the ellipsoid rounder and thereby increases the PR. As $\mu$ decreases even further, most units operate in the saturated regions of the sigmoid function, leading to a drop in the PR.

For the sub-linear network, the self-generated input tends to be much larger than in the sigmoid network because of the unbounded activation function. Thus, $\mu$ must be close to $1$ in order for the self-generated input to not push units into the saturated region of $\phi(x)$, eliminating their contribution to the PR. By contrast, fluctuations along the high-variance directions are only weakly attenuated because $\phi'(x)$ decays much more slowly with increasing $x$ than it does for the sigmoid function. Therefore, as $\mu$ decreases from $1$, the drop in the PR typically occurs before the fluctuation attenuation can produce an increase in the PR. Consistent with this, we could not find parameter combinations that produce nonmonotonic PR in sub-linear networks, although a rigorous proof remains for future work.

{\it Discussion ---}
In this letter, we showed that local connectivity balance has markedly different effects on the collective dynamics of random networks depending on the activation function. We explained these seemingly opposite effects within a unifying picture: local connectivity balance removes the self-generated feedback input. This feedback vanishes for odd activation functions because the population-averaged activity is then zero, thus the dynamical role of local connectivity balance has been masked by the common assumption of an odd activation function. In neuroscience and many other fields, however, this assumption is unrealistic, and the relevant activation functions are typically non-odd~\cite{ahmadian2021}. The effects we identify can therefore come into play in a broad range of realistic disordered networks.

We also found that local connectivity balance strongly influences the effective dimension of the collective dynamics, with the PR depending nonmonotonically on the degree of balance in sigmoid networks. Large-scale functional-connectomic datasets~\cite{bae2025}, which combine synapse-level anatomical information with recordings of neural activity, could be used to approximate local connectivity balance from the numbers and sizes of excitatory and inhibitory synapses and relate it to the dimensionality of population activity. Although synapse size and number may provide a useful proxy for local synaptic strength, the effective influence of a synapse on neuronal output also depends on its subcellular location~\cite{williams2002}, the cell types involved~\cite{campagnola2022}, and dendritic integration~\cite{london2005}. Anatomically inferred synaptic balance may therefore not precisely correspond to balance in the effective coupling strengths represented by our model; nevertheless, such an analysis could provide an indirect test of our prediction.

{\it Data availability}: The source code used to generate all figures is available at \cite{source_code}.


\begin{acknowledgments}
We thank David Clark for helpful comments on this work and Rainer Engelken for fruitful discussions on effective dimension. A.K. was supported by a Pew Biomedical Scholars Fellowship and a McKnight Scholars Award. R.G. was supported by the National Institute of Neural Disorders and Stroke under Award Number RF1NS132912.
\end{acknowledgments}


\bibliography{library}
\end{document}


\title{Supplemental Material for\\
``Local connectivity balance shapes population dynamics in random recurrent networks''}

\author{Shotaro Takasu}
\email{stakasu@scripps.edu}
\author{Richard Gast}
\author{Ann Kennedy}
\affiliation{Department of Neuroscience, Scripps Research, San Diego, CA}

\date{\today}

\maketitle
\tableofcontents

\section{Proof of no outlier eigenvalues in $\mathcal{J}$} \label{supp:no-outliers}
We show that our connectivity matrix $\mathcal{J}_{ij} = J_{ij}-\frac{\mu}{N}\sum_k^N J_{ik}$ has no outlier eigenvalues outside the bulk for any $\mu \in [0,1]$. An eigenvalue $\lambda$ of $\mathcal{J}$ satisfies $\det |\mathcal{J}-\lambda I|=0$. Since $\mathcal{J}$ is a rank-one perturbation of $J$, the matrix determinant lemma yields
\begin{align}
    \det|\lambda I - \mathcal{J}| &= \det |(\lambda - J) +\frac{\mu}{N}J \ones \ones^\top| \nonumber \\
    &= \det |\lambda - J |\left( 1 + \frac{\mu}{N} \ones^\top (\lambda -J)^{-1} J \ones \right),
\end{align}
where $\ones := (1,1,\cdots,1)^\top$ denotes the all-ones vector. For $\lambda$ outside the spectrum of $J$, the matrix $\lambda-J$ is invertible, and a possible outlier must satisfy
\begin{align} \label{eq:condition for lambda to satisfy}
    1 + \frac{\mu}{N} \ones^\top (\lambda - J)^{-1} J \ones = 0.
\end{align}
Using a Neumann series, the second term can be expanded as
\begin{align}
    \frac{\mu}{N} \ones^\top (\lambda - J )^{-1} J \ones =
    \frac{\mu}{N}\ones^\top \left( \sum_{n=1}^\infty \frac{1}{\lambda^{n}} J^n\right) \ones.
\end{align}
One can show that for any $n\geq 1$, both the mean and variance of $\frac{1}{N} \ones^\top J^n \ones$ are $O(1/N)$, so that each term vanishes in the limit $N\to \infty$. Therefore Eq.~(\ref{eq:condition for lambda to satisfy}) does not hold for any $\mu$; in other words, $\mathcal{J}$ has no outlier eigenvalues.

\section{Dynamical mean-field theory} 
\subsection{Derivation of the dynamical mean-field equation} \label{supp:dmft-derivation}
We derive the dynamical mean-field equations presented in the main text using the path-integral formalism~\cite{dahmen2020, crisanti2018}. For later use, we incorporate an external input, $\tilde{j}_i(t)$, into our network as
\begin{align}
    (1+\partial_t) x_i = \sum_j^N \mathcal{J}_{ij}(\mu) \phi_j - \tilde{j}_i.
\end{align}
Throughout this appendix, we use the shorthand notation $\bm{a}^\top \bm{b} := \sum_i^N \int dt\ a_i(t) b_i(t)$ for inner products in the field index and time, and $\bm{a}^\top B\, \bm{c} := \sum_{i,j}^N \int dt\, ds\ a_i(t) B_{ij}(t,s) c_j(s)$ for the corresponding bilinear forms. For an equal-time coupling matrix such as $\mathcal{J}$, we write $\bm{a}^\top \mathcal{J}\, \bm{c} := \sum_{i,j}^N \int dt\ a_i(t) \mathcal{J}_{ij} c_j(t)$. The generating functional of this system is defined as
\begin{widetext}
    \begin{align}
        Z[\bm{j},\bm{\tilde{j}}|\mathcal{J}] &:= \int \mathcal{D}\bm{x}\  \prod_i^N \delta\left[(1+\partial_t )x_i- \sum_j^N \mathcal{J}_{ij }\phi_j + \tilde{j}_i \right] \exp\left( \bm{j}^\top \bm{x} \right) \nonumber \\
        &= \int \mathcal{D}\bm{x} \mathcal{D}\bm{\tilde{x}}\ \exp\left( 
             \bm{\tilde{x}}^\top (1+\partial_t )\bm{x} - \bm{\tilde{x}}^\top \mathcal{J}\, \bm{\phi} + \bm{j}^\top \bm{x} + \bm{\tilde{j}}^\top \bm{\tilde{x}}
        \right),
    \end{align}
    where we have used the Fourier representation of the Dirac delta functional. Here, $\mathcal{D} \bm{x}$ and $\mathcal{D}\bm{\tilde{x}}$ are the path-integral measures. In order to take the average of the generating functional over the quenched disorder, $\mathcal{J}$, we compute $\left\langle \exp\left( - \bm{\tilde{x}}^\top \mathcal{J}\, \bm{\phi} \right) \right\rangle_\mathcal{J}$, which is given by
    \begin{align}
        \left\langle \exp\left( - \bm{\tilde{x}}^\top \mathcal{J}\, \bm{\phi} \right) \right\rangle_\mathcal{J} &= \left\langle \exp\left( -\sum_{i,j}  (J_{ij} - \frac{\mu}{N}\sum_k J_{ik}) \int dt\ \tilde{x}_i(t) \phi_j(t) \right) \right\rangle_{J} \nonumber \\
        &= \prod_{i,j} \left\langle \exp \left( - J_{ij} \int dt\ \tilde{x}_i(t)\left( \phi_j(t)- \frac{\mu}{N}\sum_k \phi_k(t) \right) \right) \right\rangle_{J} \nonumber \\
        &= \exp\left\{ \frac{1}{2} \sum_i \int dt\, ds\ \tilde{x}_i(t) 
            \underbrace{\left( \frac{g^2}{N}\sum_j (\phi_j(t)- \underbrace{\frac{\mu}{N}\sum_k \phi_k(t))}_{=: \mu m(t)}(\phi_j(s)- \underbrace{\frac{\mu}{N}\sum_k \phi_k(s)}_{=: \mu m(s)})\right)}_{=: C(t,s)} 
            \tilde{x}_i(s) 
        \right\},
    \end{align}
    where we define two order parameters: $m(t)=\frac{1}{N}\sum_k^N \phi_k(t)$ and $C(t,s)= \frac{g^2}{N} \sum_j^N (\phi_j(t)-\mu m(t)) (\phi_j(s)-\mu m(s))$. Using this expression, we obtain the quenched-averaged generating functional as
    \begin{align}
    Z[\bm{j},\bm{\tilde{j}}]
    &= \int \mathcal{D}\bm{x} \mathcal{D}\bm{\tilde{x}}\ \exp\left( 
             \bm{\tilde{x}}^\top (1+\partial_t )\bm{x} + \frac{1}{2}\, \bm{\tilde{x}}^\top C\, \bm{\tilde{x}} + \bm{j}^\top \bm{x} + \bm{\tilde{j}}^\top \bm{\tilde{x}}
        \right) \nonumber \\
    &\hspace{20pt}\times\int \mathcal{D}C \mathcal{D}m\ \delta\left[-NC(t,s) + g^2\sum_j (\phi_j(t)- \mu m(t))(\phi_j(s)- \mu m(s) )\right] \delta\left[ -Nm(t) + \sum_k \phi_k(t)\right],
    \end{align}
    where the definitions of the order parameters are imposed by inserting Dirac delta functionals. 

    Here, it is important to note that all moments of the auxiliary field, $\tilde{x}$, vanish,
    \begin{align}
        \langle \tilde{x}_{i_1}(t_1) \cdots \tilde{x}_{i_n}(t_n) \rangle_{\mathcal{J}} = \left. \frac{\partial^n}{\partial \tilde{j}_{i_1}(t_1) \cdots \tilde{j}_{i_n}(t_n)}  Z[\bm{j},\bm{\tilde{j}}]\right|_{\bm{j}=\bm{\tilde{j}}=0} = 0,
    \end{align}
    since $Z[\bm{0}, \bm{\tilde{j}}]= \langle \int \mathcal{D}\bm{x}\ p[\bm{x}|\mathcal{J}] \rangle_{\mathcal{J}} = 1$ holds independently of $\tilde{\bm{j}}$.

    Using the Fourier representation of the Dirac delta functional, the generating functional can be rewritten as
    \begin{align}
        Z[k_C,\tilde{k}_C,k_m,\tilde{k}_m] &= \int \mathcal{D}\{ C, \tilde{C},m, \tilde{m} \} 
        \exp(-N\tilde{C}^\top C - N\tilde{m}^\top m + k_C^\top C + \tilde{k}_C^\top \tilde{C} + k_m^\top m + \tilde{k}_m^\top \tilde{m} ) \\
        &\hspace{10pt}\times \int \mathcal{D}\bm{x} \mathcal{D}\bm{\tilde{x}}\ \ \exp\left\{
            \bm{\tilde{x}}^\top (1+\partial_t )\bm{x} + \frac{1}{2}\, \bm{\tilde{x}}^\top C\, \bm{\tilde{x}}
        \right. \nonumber \\
        &\hspace{60pt}\left. + g^2\sum_j \int dt\, ds\ (\phi_j(t)-\mu m(t))\tilde{C}(t,s)(\phi_j(s)-\mu m(s)) + \sum_k \int dt\ \phi_k(t) \tilde{m}(t) \right\} \nonumber \\
        &= \int \mathcal{D}\{ C, \tilde{C},m, \tilde{m} \} 
        \exp(-N\tilde{C}^\top C - N\tilde{m}^\top m + k_C^\top C + \tilde{k}_C^\top \tilde{C} + k_m^\top m + \tilde{k}_m^\top \tilde{m} ) \nonumber \\
        &\hspace{10pt}\times \left[ \int \mathcal{D}x \mathcal{D} \tilde{x}\ \exp\left\{
            \tilde{x}^\top (1+\partial_t )x +\frac{1}{2}\, \tilde{x}^\top C\, \tilde{x}+  g^2\int dt\, ds\ (\phi(t)- \mu m(t))\tilde{C}(t,s)(\phi(s)- \mu m(s)) 
        \right. \right.\nonumber \\
        &\hspace{260pt} \left. \left. + \int dt\ \phi(t) \tilde{m}(t) \right\} \right]^N \nonumber \\
        &= \int \mathcal{D}\{ C, \tilde{C},m, \tilde{m} \} 
        \exp(-N\tilde{C}^\top C - N\tilde{m}^\top m + k_C^\top C + \tilde{k}_C^\top \tilde{C} + k_m^\top m + \tilde{k}_m^\top \tilde{m} + N \mathcal{W}[C,\tilde{C},m,\tilde{m}]), \nonumber
    \end{align}
    where 
    {\small
    \begin{align}
        \mathcal{W}[C,\tilde{C},m,\tilde{m}] := \ln \int \mathcal{D}x \mathcal{D} \tilde{x}\ \exp\left\{
            \tilde{x}^\top (1+\partial_t )x +\frac{1}{2}\, \tilde{x}^\top C\, \tilde{x}+ g^2\int dt\, ds\ (\phi(t)-\mu m(t))\tilde{C}(t,s)(\phi(s)- \mu m(s)) +  \int dt\ \phi(t) \tilde{m}(t)  \right\}.
    \end{align}
    }
    \end{widetext}
    
    Here, $k_C(t,s),\tilde{k}_C(t,s), k_m(t), \tilde{k}_m(t)$ are the source fields conjugate to the order parameters $C(t,s), \tilde{C}(t,s), m(t), \tilde{m}(t)$, respectively, introduced so that derivatives of the generating functional with respect to them yield the corresponding moments of the order parameters.
    
    Using the saddle-point approximation in the limit $N\to \infty$, we obtain the following saddle-point conditions for the order parameters:
    \begin{align}
        C(t,s) &= g^2\left\langle (\phi(t)-\mu m(t))(\phi(s)- \mu m(s)) \right\rangle_\mathcal{W} \\
        m(t) &= \left\langle \phi(t) \right\rangle_\mathcal{W} \\
        \tilde{C}(t,s) &= \frac{1}{2} \left\langle \tilde{x}(t) \tilde{x}(s) \right\rangle_\mathcal{W} \\
        \tilde{m}(t) &= 2 \mu g^2\left\langle \int ds\  \tilde{C}(t,s) (\mu m(s)-\phi(s)) \right\rangle_\mathcal{W} .
    \end{align}
    As noted above, the moments of the auxiliary fields vanish, which leads to $\tilde{C}=\tilde{m}=0$. Therefore, the generating functional reduces to
    \begin{align}
        &Z[k_C,\tilde{k}_C,k_m,\tilde{k}_m] = \int \mathcal{D}C \mathcal{D}m\  
        \exp(  k_C^\top C  + k_m^\top m + N \mathcal{W}[C,m]) \\
        &\mathcal{W}[C,m] = \ln \int \mathcal{D}x \mathcal{D} \tilde{x}\ \exp\left(
            \tilde{x}^\top (1+\partial_t )x +\frac{1}{2}\, \tilde{x}^\top C\, \tilde{x} \right).
    \end{align}
    This is exactly equivalent to a system of $N$ mutually independent units, each evolving according to
    \begin{align}
        &(1+\partial_t) x(t) = \eta(t),\hspace{20pt} \eta \sim GP(0, C) \\
        &C(t,s) = g^2 \left\{ \langle \phi(t) \phi(s) \rangle_\eta  - \mu(2-\mu) \langle \phi(t) \rangle_\eta \langle \phi(s) \rangle_\eta \right\} \nonumber \\
        &\hspace{30pt}=  g^2 \left\{ \llangle \phi(t),\phi(s) \rrangle_\eta +(1-\mu)^2 \langle \phi(t) \rangle_\eta \langle \phi(s) \rangle_\eta \right\}.
    \end{align}

\subsection{Phase transition point in saturating and sublinear networks} \label{supp:phase-transition-saturating-sublinear}

Here, we analytically show that in networks with saturating or sub-linearly scaling activation functions, the phase transition point $g_c$ from fixed-point dynamics to chaos decreases as $\mu$ increases. As mentioned in the main text, we make the following natural assumptions on the activation function: (1) it is increasing, (2) $\lim_{x\to\pm\infty} \phi'(x)=0$, and (3) $\phi'(x)$ is unimodal with its peak at $x=0$. The second condition ensures that $\phi$ scales sub-linearly or more slowly, which prevents divergence and allows chaotic dynamics.

\begin{figure}[t]
    \centering
    \includegraphics[width=0.7\textwidth]{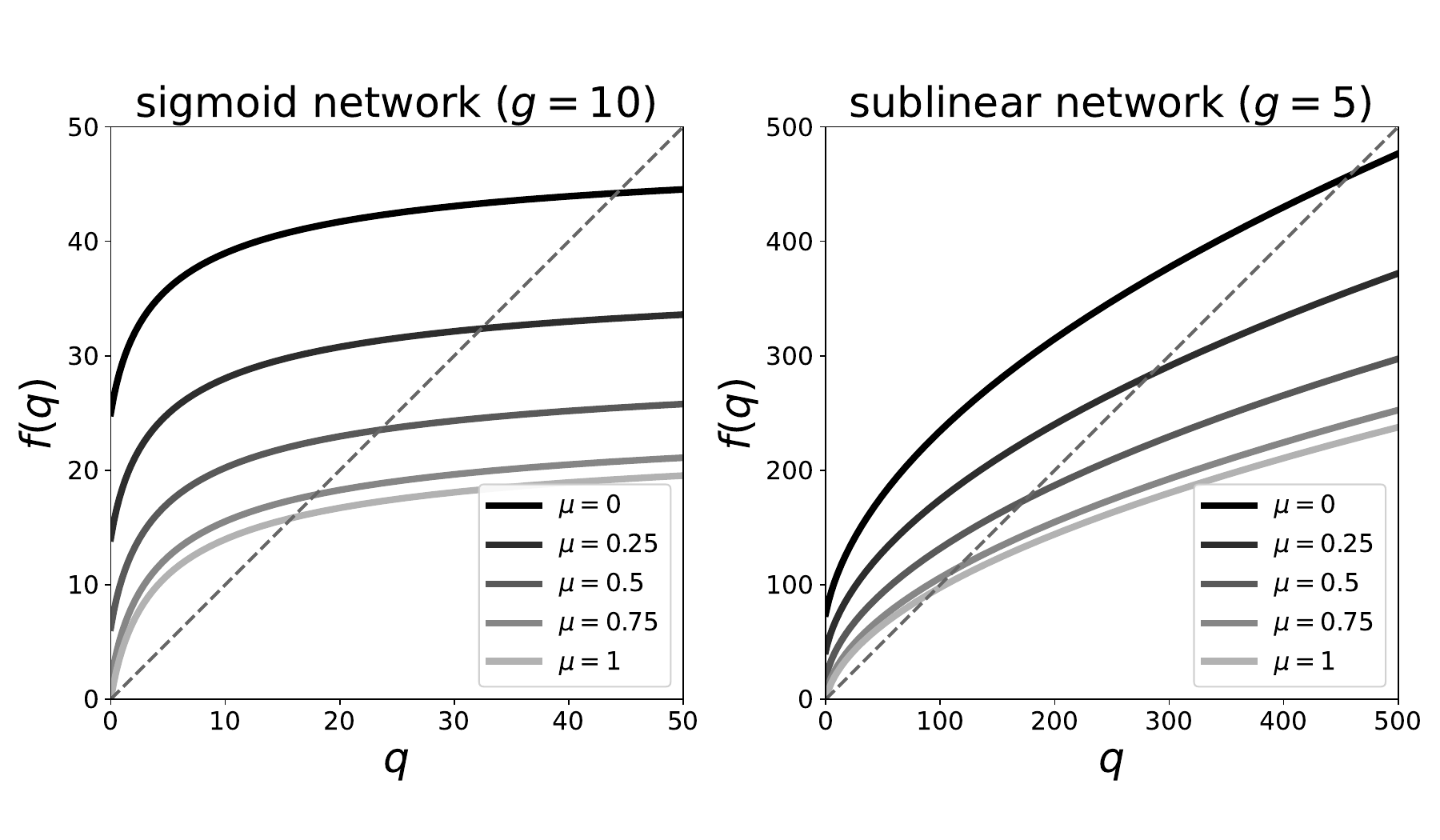}
    \caption{
        Graphical solutions of the self-consistent equation $q=f(q)$ for the sigmoid activation function $\phi(x)=1/(1+e^{-x})$ (left) and the sub-linearly scaling function $\phi(x)=\sqrt{(x+1)+\sqrt{(x+1)^2+3}}$ (right). The intersections of the curves with the diagonal give the solutions $q^*$.
    }
    \label{fig:self-consistent eq}
\end{figure}

Assuming that the network state is at a fixed point, the preactivations $\{ x_i^0 \}_{i=1}^N$ follow $\mathcal{N}(0, q^*)$, where the variance $q^*$ is determined by the self-consistent equation $q=f(q)$, with
\begin{align} \label{eq:supp self-consistent equation for fixed point}
    f(q) :=g^2 \left\{  \llangle \phi,\phi \rrangle_{x\sim\mathcal{N}(0,q)} +(1-\mu)^2 \langle \phi \rangle_{x\sim\mathcal{N}(0,q)}^2\right\}.
\end{align}
Because of the assumptions on $\phi$, $f(q)$ is generically an increasing, concave-down function. Therefore, the equation $q=f(q)$ has a unique solution $q=q^*$ as long as $f(0)>0$. When $\mu=1$ or $\phi$ is an odd function, $f(0)=0$ holds. In this case, whether a nonzero solution exists depends on whether the slope of $f$ at the origin, $ f'(0)=g^2\phi'(0)^2$, exceeds one: $q=0$ for $g\phi'(0) < 1$, and $q=q^*>0$ for $g\phi'(0) > 1$. Because $f(q)$ is a decreasing function of $\mu$, the nonzero solution $q^*$ decreases as $\mu$ increases, as can be seen graphically in Fig.~\ref{fig:self-consistent eq}.

Next, we examine the stability of the fixed point $x_i^0 \sim \mathcal{N}(0, q^*)$. It becomes linearly unstable once the spectral radius $\rho$ of the Jacobian matrix $\mathcal{J}\phi'(\bm{x^0})$ exceeds one, where $\rho = g\sqrt{\langle \phi'(x^0)^2 \rangle_{x^0 \sim \mathcal{N}(0, q^*)}}$~\cite{ahmadian2015}. Hence the phase transition point is given by $g_c = 1/\sqrt{\langle \phi'(x^0)^2 \rangle_{x^0 \sim \mathcal{N}(0, q^*)}}$. Given the third assumption on $\phi$, $\langle \phi'(x^0)^2 \rangle_{x^0}$ increases as $q^*$ decreases. Since $q^*$ decreases with increasing $\mu$, the spectral radius grows and $g_c$ correspondingly decreases. Taken together, we conclude that the transition point $g_c$ is a decreasing function of $\mu$.

\subsection{Phase transition point in ReLU networks} \label{supp:phase-transition-relu}

Here, we analytically derive the critical coupling strength $g_c$ separating the regime with a stable zero fixed point from the divergent regime in ReLU networks. Starting from the dynamical mean-field equations derived in Sec.~\ref{supp:dmft-derivation}, we obtain
\begin{align}
    (1+\partial_t)(1+\partial_s)\Delta(t,s) = C(t,s),
\end{align}
where $\Delta(t,s):=\langle x(t)x(s)\rangle_\eta$. Assuming stationarity, this equation can be rewritten as 
\begin{align} \label{eq:dmft for stationary state}
    (1-\partial_\tau^2)\Delta(\tau) = C(\tau),
\end{align}
where $\Delta(\tau):=\Delta(t,t+\tau)$ and $C(\tau):=C(t,t+\tau)$~\cite{sompolinsky1988, dahmen2020}.

To examine stability near $\bm{x}=\bm{0}$, we write $\Delta(\tau)=\epsilon R(\tau)$, where $\epsilon=\Delta(0)\ll 1$ and $R(0)=1$. For the ReLU activation function, the Gaussian averages entering $C(\tau)$ can be evaluated analytically, yielding
\begin{align} \label{eq:C(tau) as a function of R(tau)}
    C(\tau) = \epsilon \frac{g_c^2}{2\pi} \Big[ 
        \sqrt{1-R(\tau)^2} + R(\tau)\big(\pi- \arccos R(\tau)\big) 
         -\mu(2-\mu) \Big]
\end{align}
Importantly, $C(\tau)$ is proportional to $\epsilon$, and hence has the same overall scale as $\Delta(\tau)$. This follows from the scale invariance of the ReLU function, $\phi(ax)=a\phi(x)$ for $a\geq 0$. Consequently, $\epsilon$ cancels from the dynamical mean-field equation, leaving the overall amplitude undetermined. We may therefore set $\Delta(0)=1$ without loss of generality and, for notational simplicity, relabel $R$ as $\Delta$. Eqs.~(\ref{eq:dmft for stationary state}) and~(\ref{eq:C(tau) as a function of R(tau)}) then give
\begin{align} \label{eq:dmft for relu}
    (1-\partial_\tau^2) \Delta(\tau) = g_c^2 h(\Delta(\tau),\mu)
\end{align}
where 
\begin{align}
    h(\Delta,\mu) := \frac{1}{2\pi} \Big[ \sqrt{1-\Delta^2}+\Delta \big( \pi- \arccos \Delta \big)-\mu(2-\mu) \Big].
\end{align}
The boundary conditions are 
\begin{align} \label{eq:boundary condition Appendix}
    \Delta(0)=1,\ \Delta'(0)=0,\ \lim_{\tau \to \infty}\Delta'(\tau)=0.
\end{align}
Here, the second condition follows from the symmetry $\Delta(\tau)=\Delta(-\tau)$, whereas the third assumes that the autocovariance approaches a finite plateau at long time lags.

Taking $\tau\to\infty$ in Eq.~(\ref{eq:dmft for relu}) gives
\begin{align} \label{eq:infinite condition}
    \Delta(\infty)=g_c^2 h(\Delta(\infty), \mu).
\end{align}
Eq.~(\ref{eq:dmft for relu}) can also be interpreted as the equation of motion of a particle in a potential $U(\Delta)$~\cite{sompolinsky1988,dahmen2020}:
\begin{align}
    \partial_\tau^2 \Delta &= -\partial_\Delta U(\Delta) \\
    U(\Delta) &= -\frac{1}{2}\Delta^2 + g_c^2 \int_{0}^{\Delta} h(\Delta, \mu) d\Delta.
\end{align}
Here the position of the particle at time $\tau$ is identified with $\Delta(\tau)$, and its initial and asymptotic conditions are given by Eq.~(\ref{eq:boundary condition Appendix}). Applying energy conservation between $\tau=0$ and $\tau \to \infty$ yields 
\begin{align} \label{eq:energy conservation}
    1-\Delta(\infty)^2 = 2 g_c^2 \int_{\Delta(\infty)}^1 h(\Delta,\mu) d\Delta.
\end{align}
Combining Eqs.~(\ref{eq:infinite condition}) and (\ref{eq:energy conservation}) gives
\begin{align}
    g_c = \left[ \frac{2}{1-\Delta(\infty)^2} \int_{\Delta(\infty)}^{1} h(\Delta, \mu) d\Delta \right]^{-\frac{1}{2}},
\end{align}
where $\Delta(\infty)$ is given implicitly by
\begin{align}
   (1-\Delta(\infty)^2)h(\Delta(\infty), \mu) = 2\Delta(\infty)  \int_{\Delta(\infty)}^{1} h(\Delta, \mu) d\Delta.
\end{align} 

\subsection{Vanishing time-averaged preactivation at $\mu=1$ in the large-N limit} \label{supp:zero-time-mean}

Here, we prove that when $\mu =1$, all units satisfy $\langle x_i \rangle_t =0$. The dynamical mean-field equation for $\mu=1$ is
\begin{align}
    &(1+\partial_t) x(t) = \eta(t) \\
    &\eta \sim GP(0, C) \\
    &x \sim GP(0, \Delta) \\
    &C(\tau) = g^2 \langle \phi(x(t))\phi(x(t+\tau)) \rangle_{x} - g^2\langle \phi(x) \rangle_x^2.
\end{align}
The population average of $\langle x_i \rangle_t$ is zero because $x$ is a Gaussian process with mean $0$. Let $V$ be the population variance of $\langle x_i \rangle_t$. Then, it is enough to show $V=0$.

\subsubsection{Fixed-point regime}

When the network state is in the fixed-point regime, it directly follows that $C=\Delta=V$ holds. The self-consistent equation for $V$ is thus 
\begin{align}
    V &= g^2 \langle \phi(x)^2 \rangle_{x \sim \mathcal{N}(0, V)} - g^2 \langle \phi(x) \rangle^2_{x \sim \mathcal{N}(0, V)} 
    = g^2 \mathbb{V}_{x \sim \mathcal{N}(0, V)}[\phi(x)].
\end{align}
The trivial solution is $V=0$. Using the Gaussian--Poincar\'e inequality, 
\begin{align}
    g^2 \mathbb{V}_{x \sim \mathcal{N}(0, V)}[\phi(x)] &\leq g^2V \langle \phi'(x)^2 \rangle_{x \sim \mathcal{N}(0, V)} < g^2 \phi'(0)^2 V,
\end{align}
where the second inequality follows from the assumption that $\phi'(x)$ is unimodal with a peak at $x=0$. Since the network is in the fixed-point regime, $g$ satisfies $g<1/\phi'(0)$. Therefore, we obtain $g^2 \mathbb{V}_{x \sim \mathcal{N}(0, V)}[\phi(x)] < V$, which means that $V=0$ is the only solution.

\subsubsection{Chaos regime}
First, we show that $\langle x_i \rangle_t =0\ (\forall i)$ holds if and only if $\lim_{\tau \to \infty}\Delta(\tau) =0$. By definition, we have
\begin{align}
    V &= \left\langle  \left( \lim_{T \to \infty} \frac{1}{T}\int_0^T x_i(t) dt \right)^2 \right\rangle_x 
    = \lim_{T\to \infty} \frac{1}{T^2} \int_0^T \int_0^T  \Delta(s,t)\  ds dt 
    = \lim_{T\to \infty} \frac{1}{T} \int_{-T}^T \Delta(\tau) \left(1-\frac{|\tau|}{T} \right) d\tau,
\end{align}
where we have assumed time-translation symmetry, i.e., $\Delta(t, t+\tau) =: \Delta(\tau)$ is independent of $t$, which holds when the system is in a stationary state. We decompose $\Delta(\tau)$ as $\Delta(\tau) = \Delta_\infty + \Delta_{\rm rs} (\tau)$, where $\Delta_\infty := \lim_{\tau \to \infty} \Delta(\tau)$. Substituting this into the equation above yields
\begin{align}
    V = \Delta_\infty + \lim_{T\to \infty} \frac{2}{T} \int_{0}^T \Delta_{\rm rs}(\tau) \left(1-\frac{\tau}{T} \right) d\tau.
\end{align}
The second term vanishes as
\begin{align}
    \left| \frac{2}{T} \int_{0}^T \Delta_{\rm rs}(\tau) \left(1-\frac{\tau}{T} \right) d\tau \right| \leq \frac{2}{T} \int_{0}^T |\Delta_{\rm rs}(\tau)| \left| 1-\frac{\tau}{T} \right| d\tau 
    \leq \frac{2}{T} \int_{0}^T |\Delta_{\rm rs}(\tau)|  d\tau 
    \overset{T\to\infty}{\longrightarrow} 0.
\end{align}
Here, we assumed that $\Delta_{\rm rs}(\tau)$ is integrable. As a result, we obtain $V=\Delta_\infty$, which shows that $\langle x_i \rangle_t =0\ (\forall i)$, i.e., $V=0$, holds if and only if $\lim_{\tau \to \infty}\Delta(\tau) = 0$.

Next, we show $\lim_{\tau \to \infty}\Delta(\tau) =0$ for $\mu=1$. Following the same procedure as in Sec.~\ref{supp:phase-transition-relu}, we obtain from the dynamical mean-field equation
\begin{align}
    &(1-\partial_\tau^2) \Delta(\tau) = f(\Delta(\tau); \Delta_0) \\
    &f(\Delta(\tau); \Delta_0) := g^2 \langle \phi(x) \phi(y) \rangle_{(x,y) \sim \mathcal{N}(0, \Sigma(\tau))} - g^2 \langle \phi(x) \rangle_{x \sim \mathcal{N}(0, \Delta_0 ) }^2 \\
    &\Sigma(\tau) = \begin{pmatrix} \Delta_0 & \Delta(\tau) \\ \Delta(\tau) & \Delta_0 \end{pmatrix}.
\end{align}
This equation can also be regarded as an equation of motion: 
\begin{align}
    &\partial_\tau^2 \Delta = - \partial_\Delta U(\Delta; \Delta_0) \\
    &U(\Delta(\tau); \Delta_0) = - \frac{1}{2} \Delta(\tau)^2 +\int_{0}^{\Delta(\tau)}  f(\Delta; \Delta_0) d\Delta, \label{eq:potential}
\end{align}
where $U(\Delta; \Delta_0)$ is a potential function of $\Delta$ conditioned on $\Delta_0$. Here the position of the particle at time $\tau$ is identified with $\Delta(\tau)$, and the initial conditions are $\Delta(0)=\Delta_0$ and $\Delta'(0)=0$.

To gain intuition about the solution $\Delta(\tau)$, it is helpful to examine the shape of $U$. One readily sees that $U(0;\Delta_0)=0$ and $\partial_\Delta U(\Delta; \Delta_0)|_{\Delta = 0} = 0$. The gradient of $U$ is
\begin{align}
    \partial_{\Delta} U(\Delta; \Delta_0) = - \Delta + f(\Delta; \Delta_0).
\end{align}
The function $f(\Delta; \Delta_0)$ is increasing and convex for $\Delta \geq 0$, because using Price's theorem, we can evaluate $\partial_\Delta f$ and $\partial_\Delta^2 f$ as 
\begin{align}
    \partial_\Delta f &= g^2 \langle \phi'(x) \phi'(y) \rangle_{(x,y) \sim \mathcal{N}(0, \Sigma(\tau))} \geq 0\ \  (\because \phi'(x) \geq 0) \\ 
    \partial_\Delta^2 f &= g^2 \langle \phi''(x) \phi''(y) \rangle_{(x,y) \sim \mathcal{N}(0, \Sigma(\tau))} \nonumber \\
    &= g^2 \mathbb{E}_{z} \left[ \mathbb{E}_{u}[\phi''(\sqrt{\Delta}z + \sqrt{\Delta_0-\Delta}u)]  \times \mathbb{E}_{v}[\phi''(\sqrt{\Delta}z + \sqrt{\Delta_0-\Delta}v)] \right]\ (\because \Delta \geq 0) \nonumber \\
    & =  g^2 \mathbb{E}_{z} \left[ \left\{ \mathbb{E}_{u}[\phi''(\sqrt{\Delta}z + \sqrt{\Delta_0-\Delta}u)] \right\}^2\right] \geq 0.
\end{align}

\begin{figure}[tbp]
    \centering
    \includegraphics[width=0.7\textwidth]{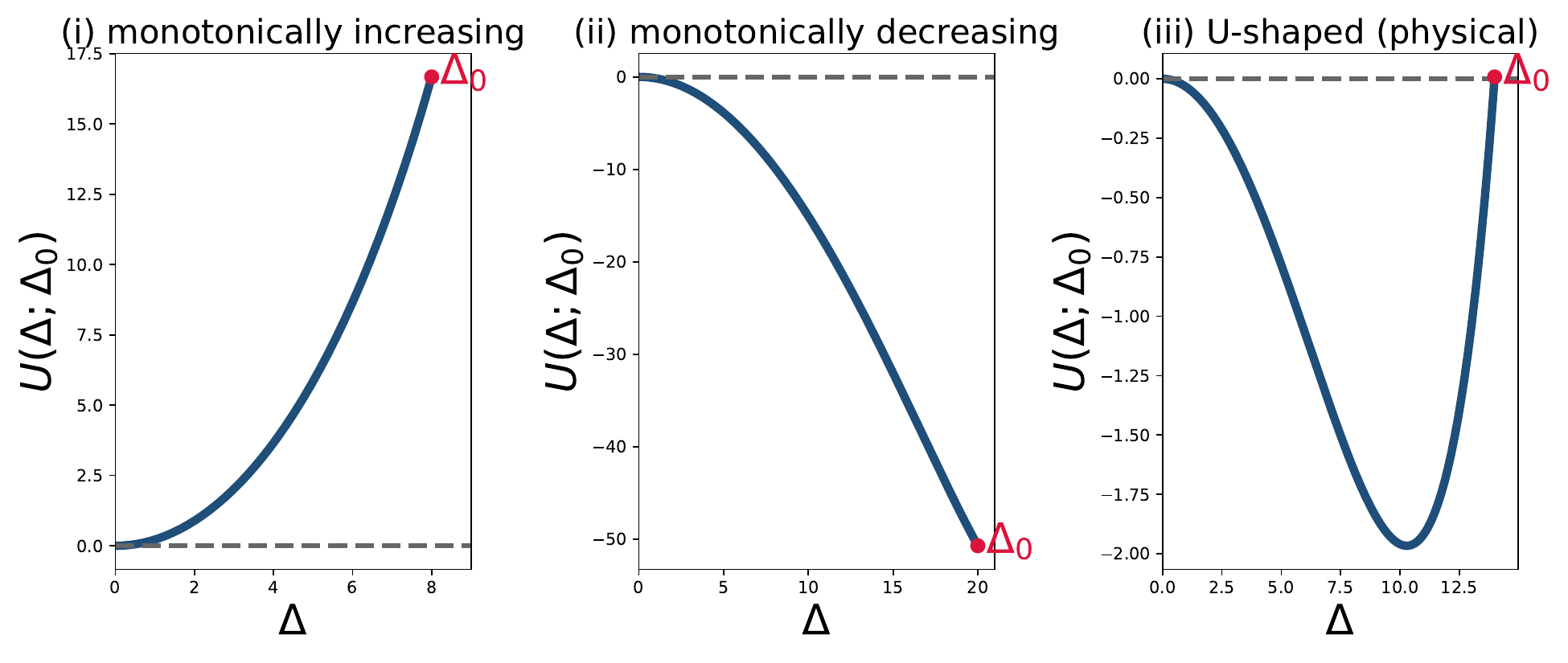}
    \caption{%
    The three possible shapes of the potential function $U(\Delta;\Delta_0)$, computed for the sigmoid activation function with $g=10$. The value of $\Delta_0$ is varied: $\Delta_0=8.0$ for ({\romannumeral 1}), $20.0$ for ({\romannumeral 2}), and $13.98$ for ({\romannumeral 3}). The potential $U$ is obtained by numerically integrating Eq.~(\ref{eq:potential}).
    }
    \label{fig:potential shapes}
\end{figure}

Therefore, $U(\Delta; \Delta_0)$ can take three possible shapes depending on the values of $g$ and $\Delta_0$ (Fig.~\ref{fig:potential shapes}): ({\romannumeral 1}) a monotonically increasing function when $f'(0)= g^2 \langle \phi'(x) \rangle^2_{x \sim \mathcal{N}(0,\Delta_0)}>1$; ({\romannumeral 2}) a monotonically decreasing function when $\Delta_0 > f(\Delta_0; \Delta_0)$; and ({\romannumeral 3}) a function with a minimum at $\Delta = \Delta^*$, satisfying $\Delta^*=f(\Delta^*)$, when $f'(0)=g^2 \langle \phi'(x) \rangle^2_{x \sim \mathcal{N}(0,\Delta_0)}<1$ and $\Delta_0 < f(\Delta_0; \Delta_0)$.

Let us consider each scenario. ({\romannumeral 1}) If $U$ were monotonically increasing, the particle would pass through $\Delta=0$, leading to $\lim_{\tau \to \infty} \Delta(\tau) <0$. This contradicts $\lim_{\tau \to \infty} \Delta(\tau) = V \geq 0$, so this solution is unphysical.

({\romannumeral 2}) If $U$ were monotonically decreasing, the particle would be driven toward larger $\Delta$, i.e., $\Delta(\tau)>\Delta_0$, which is prohibited by $|\Delta(\tau)| \leq \Delta_0$. This solution is therefore unphysical as well.

({\romannumeral 3}) The U-shaped potential gives the physical solution. The particle is released from $\Delta_0 > 0$ with $\partial_\tau \Delta|_{\tau=0} = 0$, and the boundary condition $\lim_{\tau\to\infty}\Delta(\tau) = V \geq 0$ selects the separatrix trajectory that asymptotically comes to rest at the local maximum $\Delta = 0$. By energy conservation, this requires the starting point to satisfy $U(\Delta_0; \Delta_0) = 0$, which determines $\Delta_0$. Along this trajectory, the particle slides down into the well, passes the bottom $\Delta^*$, and climbs the opposite slope toward $\Delta = 0$, which it reaches only as $\tau \to \infty$. The resulting $\Delta(\tau)$ thus decays monotonically from $\Delta_0$ to zero, giving the decaying autocorrelation function characteristic of a chaotic state.

Therefore, we conclude that $\lim_{\tau \to \infty} \Delta(\tau)=0$, and since $V=\Delta_\infty$, this gives $V=0$; i.e., $\langle x_i \rangle_t =0$ holds for every unit. Importantly, since $\partial_\Delta U(\Delta;\Delta_0)|_{\Delta =0}>0$ for $\mu<1$, the point $\Delta=0$ is no longer the local maximum of $U$ in the absence of complete balance. The local maximum, which the trajectory approaches as $\tau \to \infty$, and which therefore determines $\lim_{\tau \to \infty} \Delta(\tau)$, shifts toward larger $\Delta$. Consequently, $\lim_{\tau \to \infty} \Delta(\tau) > 0$, meaning that the time-averaged preactivations $\langle x_i \rangle_t$ are distributed with mean zero and non-zero variance when $\mu<1$.

\bibliography{library}